\documentclass[aps,pra,twocolumn,amsmath,amssymb,nofootinbib,superscriptaddress]{revtex4}

\newcommand{\bra}[1]{\langle#1|}
\newcommand{\ket}[1]{|#1\rangle}

\usepackage[pdftex]{graphicx}
\usepackage{mathrsfs}

\begin{document}

\bibliographystyle{apsrev}

\title{Optical quantum computing with photons of arbitrarily low fidelity and purity}

\author{Peter P. Rohde}
\email[]{dr.rohde@gmail.com}
\homepage{http://www.peterrohde.org}
\affiliation{Centre for Engineered Quantum Systems, Department of Physics and Astronomy, Macquarie University, Sydney NSW 2113, Australia}

\date{\today}

\frenchspacing

\begin{abstract}
Linear optics quantum computing (LOQC) is a leading candidate for the implementation of large scale quantum computers. Here quantum information is encoded into the quantum states of light and computation proceeds via a linear optics network. It is well known that in such schemes there are stringent requirements on the spatio-temporal structure of photons -- they must be completely indistinguishable and of very high purity. We show that in the Boson-sampling model for LOQC these conditions may be significantly relaxed. We present evidence that by increasing the size of the system we can implement a computationally hard algorithm even if our photons have arbitrarily low fidelity and purity. These relaxed conditions make Boson-sampling LOQC within reach of present-day technology.
\end{abstract}

\maketitle

\section{Introduction}

Linear optics quantum computation (LOQC) \cite{bib:KLM01,bib:KokLovett10} has emerged as one of the leading candidates for the implementation of scalable quantum computation (QC) \cite{bib:NielsenChuang00}. Here information is encoded into single photon states, and the computation proceeds via a network of linear optics elements, complemented by measurement and feedforward. It is known that LOQC is universal for quantum computation \cite{bib:KLM01}. However, owing to fast feedforward, the required technology is challenging and well beyond the capabilities of present-day experiments. An alternate approach, known as `Boson-sampling', was recently presented by Aaronson \& Arkhipov (AA) \cite{bib:AaronsonArkhipov10}, which does away with fast feedforward, requiring only single photon states, a passive linear network and photo-detection. While not believed to be universal for quantum computation, AA presented strong evidence that such schemes implement an algorithm classically hard to simulate, making it of interest to the quantum computing community who wish to demonstrate devices with capabilities beyond classical computers.

Like any QC architecture, LOQC is plagued by difficulties. In the Boson-sampling model the dominant sources of errors are loss, detector and source inefficiency, impurity of photons, and photon distinguishability (caused by mode-mismatch or non-identical photon sources). Recently Rohde \& Ralph \cite{bib:RohdeRalphAA12} considered the issue of loss in the Boson-sampling model, and presented evidence that within realistically achievable bounds, lossy Boson-sampling computers remain classically hard to simulate. In this paper we consider the spectral structure of photons, another limiting factor in the implementation of Boson-sampling. Conventional wisdom is that LOQC places stringent demands on the fidelity and purity requirements of photons, which is very technologically challenging. We present evidence that in the setting of Boson-sampling these tough requirements may be relaxed, allowing us to implement a computationally hard algorithm even with photons of low fidelity and purity. This makes elementary demonstrations of Boson-sampling much more foreseeable than other LOQC protocols.

Closely related to Boson-sampling is the quantum walk model \cite{bib:ADZ,bib:AAKV,bib:Kempe08, bib:Salvador12}, which has recently attracted much experimental interest \cite{bib:Schreiber10,bib:Broome10,bib:Peruzzo10,bib:Schreiber11b,bib:Schreiber12}. It was argued in Ref. \cite{bib:RohdeSchreiber12} that a multi-walker photonic quantum walk is in fact equivalent to Boson-sampling. Thus, our results can be interpreted as applying to multi-walker quantum walks also.

The results we present apply only to the restricted Boson-sampling model for LOQC and are not applicable to universal LOQC schemes such as that by Knill, Laflamme \& Milburn \cite{bib:KLM01}, or recent measurement-based protocols \cite{bib:Raussendorf01, bib:Raussendorf03}.

\section{Boson-sampling}

\begin{figure}[!htb]
\includegraphics[width=0.5\columnwidth]{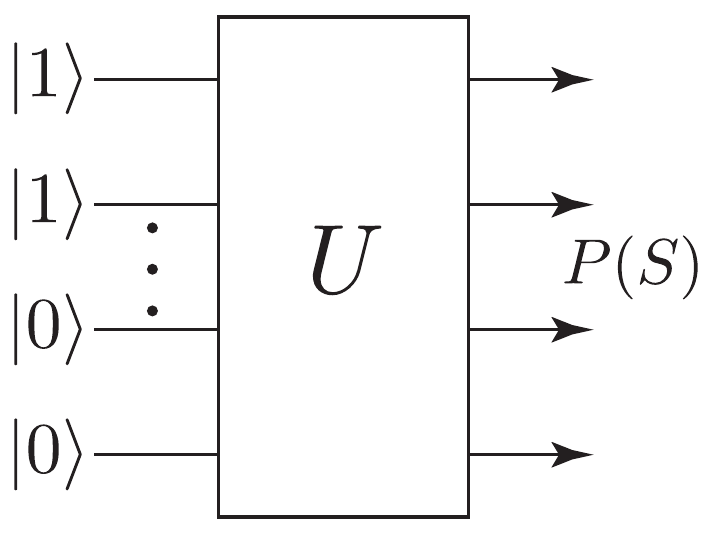}
\caption{Model for Boson-sampling linear optics quantum computation. We begin with some number of modes $m$. The first $n$ modes are initialised with a single photon state and the remainder with the vacuum state. The input state passes through a passive linear network comprising beamsplitters and phase-shifters. We repeat the experiment many times, sampling the output probability statistics.} \label{fig:model}
\end{figure}

In the Boson-sampling model we begin with an $n$ photon state across $m$ modes, of the form
\begin{equation} \label{eq:ideal_input_state}
\ket{\psi_\mathrm{in}} = \ket{1_1,\dots,1_{n},0_{n+1},\dots,0_{m}},
\end{equation}
where $m=O(n^2)$. The input state passes through a linear network, which applies a unitary map to the photon creation operators,
\begin{equation}
a_i^\dag \to \sum_{j=1}^m U_{ij} a_j^\dag,
\end{equation}
where $a_i^\dag$ is the photon creation operator associated with spatial mode $i$. In general, in an occupation number representation, the output state is of the form
\begin{equation} 
\ket{\psi_\mathrm{out}} = \sum_S \chi_S \ket{n_1^{(S)},n_2^{(S)}\dots,n_m^{(S)}},
\end{equation}
where $\sum_{i=1}^m n_i^{(S)} = n \,\forall\, S$, $S$ denotes a photon number configuration at the output, and the number of configurations \mbox{$|S| = \binom{m+n-1}{n}$} grows exponentially against the number of photons. The probability of measuring some configuration is \mbox{$P(S) = |\chi_S|^2 \propto \mathrm{Per}(A_S)$} for some matrix $A_S=f(U,S)$. Calculating matrix permanents is known to reside in the complexity class \textbf{\#P}-complete, a class strongly believed to be classically hard to solve, giving rise to the believed hardness of Boson-sampling. See Ref. \cite{bib:AaronsonArkhipov10} for a much more rigorous complexity argument. The model is illustrated in Fig. \ref{fig:model}.
We will denote an instance of a Boson-sampling computer with $n$ photons as $\mathrm{BosonSamp}(n)$.

It is known that a universal quantum computer can simulate Boson-sampling, but the converse is not believed to be the case. Specifically, it is known that \mbox{$\mathbf{BosonSampP}\subseteq\mathbf{SampBQP}$}, but strongly believed that \mbox{$\mathbf{BosonSampP}\subset\mathbf{SampBQP}$}.

\section{The mode structure of photons}

In most usual treatments, one represents Fock states as $\ket{n}$, or using some power of photon creation operators $a^\dag$. In the present study we are interested in the temporal/spectral properties of photons, thus this representation no longer suffices. Instead we follow Ref. \cite{bib:RohdeMauererSilberhorn07} and employ the mode operator formalism for representing photonic states. Here we replace the photon creation operator with a mode operator defined as
\begin{equation}
A_{\psi,j}^\dag = \int \psi(\omega) a_j^\dag(\omega)\, \mathrm{d}\omega,
\end{equation}
where $A_{\psi,j}^\dag$ is a creation operator creating a photon with spectral distribution function $\psi(\omega)$ in spatial mode $j$, $a_j^\dag(\omega)$ is a single frequency photonic creation operator in spatial mode $j$, and $\psi(\omega)$ is a normalised spectral distribution function satisfying
\begin{equation}
\int |\psi(\omega)|^2 \,\mathrm{d}\omega = 1.
\end{equation}
All our integrals are implicitly in the range $(0,\infty)$. Thus, the mode operators can be regarded as photonic creation operators that generate photons with a well defined spectral structure. Such a spectral decomposition of photonic states has been employed previously in a variety of situations \cite{bib:RohdeRalph05, bib:RohdePryde05, bib:RohdeRalph05b, bib:RohdeRalphMunro06, bib:RohdeRalph06, bib:RohdeRalph06b, bib:Rohde06b}. A similar decomposition could be employed in the time-domain, simply by taking the Fourier transform of $\psi(\omega)$, in the $x-y$ spatial degrees of freedom using a double integral, or in the polarisation degree of freedom by considering a two-element discrete sum as opposed to an integral. But we will focus just on the spectral structure of photons for ease of exposition.

Next, following Ref. \cite{bib:RohdeMauererSilberhorn07}, we can decompose the spectral distribution function into a discrete orthonormal basis of functions $\xi_i(\omega)$,
\begin{equation}
\psi(\omega) = \sum_i \lambda_i \xi_i(\omega),
\end{equation}
where the coefficients $\lambda_i$ can be calculated from
\begin{equation}
\lambda_i = \int \xi_i(\omega)^* \psi(\omega)\, \mathrm{d}\omega.
\end{equation}
For orthonormality, $\xi_i(\omega)$ satisfy
\begin{equation} \label{eq:orthonormality}
\int \xi_i(\omega)^*\xi_j(\omega)\,\mathrm{d}\omega = \bra{0} A_{\xi_i} A_{\xi_j}^\dag \ket{0} = \delta_{i,j}.
\end{equation}
Then, our mode operators can be re-expressed as
\begin{eqnarray}
A_{\psi,j}^\dag &=& \sum_i \lambda_i \int \xi_i(\omega) a_j^\dag(\omega)\, \mathrm{d}\omega \nonumber \\
&=& \sum_i \lambda_i A_{\xi_i,j}^\dag.
\end{eqnarray}
Any basis $\{\xi_i\}$ satisfying the constraint from Eq. \ref{eq:orthonormality} could be employed. Examples could include Fourier bases, wavelet bases, Hermite polynomial bases, or bases of frequency or temporal delta functions. For our study we will remain general and not restrict ourselves to any particular basis.

Using such a decomposition one can easily define metrics such as the overlap between two photonic states,
\begin{eqnarray}
\mathcal{O}_{\psi_1,\psi_2} &=& \bra{0} A_{\psi_1} A_{\psi_2}^\dag \ket{0} \nonumber \\
&=& \int \psi_1(\omega)^* \psi_2(\omega)\,\mathrm{d}\omega \nonumber \\
&=& \sum_i \lambda_{i,\psi_1}^* \lambda_{i,\psi_2},
\end{eqnarray}
and the fidelity can be defined as
\begin{equation}
\mathcal{F}_{\psi_1,\psi_2} = |\mathcal{O}_{\psi_1,\psi_2}|^2.
\end{equation}

For spectrally mixed states, in a density operator formalism we write
\begin{eqnarray}
\rho_{\psi,j} &=& \sum_i \gamma_{i,j} A_{\xi_i,j}^\dag\ket{0}\bra{0}A_{\xi_i,j} \nonumber \\
&=& \sum_i \gamma_{i,j} \rho_{\xi_i,j}.
\end{eqnarray}
That is, the state is a mixture over different distributions $\xi_i$. Note that each sub-distribution $\xi_i$ is pure. The purity of a mixed single photon state is
\begin{equation}
\mathcal{P_\psi} = \mathrm{tr}(\rho_{\psi,j}^2) = \sum_i \gamma_{i,j}^2.
\end{equation}

\section{Boson-sampling with arbitrary photons}

We now consider the effects the spectral properties of the input photons have on the operation of Boson-sampling, considering the cases of both spectrally pure and spectrally mixed photons.

\subsection{Spectrally mixed photons}

We begin by considering the effects of spectrally impure/mixed photons on the operation of Boson-sampling. Our input state can be expressed
\begin{eqnarray} \label{eq:input_mixed}
\rho_\mathrm{in} &=& \bigotimes_{j=1}^n \rho_{\psi_j,j} \nonumber \\
&=& \bigotimes_{j=1}^n \sum_i \gamma_{i,j} \rho_{\xi_i,j}.
\end{eqnarray}

Expanding this expression we obtain a mixture of different combinations of $n$ photon input states distributed across the different spectral basis functions. As a simple illustrative example, suppose our Boson-sampling computer has $n=2$ photons, $m=2$ modes, and the spectral distribution functions are supported by two basis functions $\xi_1$ and $\xi_2$. Then the expansion of the input state from Eq. \ref{eq:input_mixed} is of the form
\begin{eqnarray} \label{eq:expansion_simple}
\rho_\mathrm{in} &=& \gamma_{1,1}\gamma_{1,2} \cdot \rho_{\xi_1,1}\otimes\rho_{\xi_1,2} \nonumber \\
&+& \gamma_{1,1}\gamma_{2,2} \cdot \rho_{\xi_1,1}\otimes\rho_{\xi_2,2} \nonumber \\
&+& \gamma_{2,1}\gamma_{1,2} \cdot \rho_{\xi_2,1}\otimes\rho_{\xi_1,2} \nonumber \\
&+& \gamma_{2,1}\gamma_{2,2} \cdot \rho_{\xi_2,1}\otimes\rho_{\xi_2,2}.
\end{eqnarray}
Next, note that because $\{\xi_i\}$ form an orthonormal basis, if two photons are present, each in the same spectral basis state, they will interfere as expected in the ideal case, whereas if they are in different spectral basis states, they will not. Thus, \mbox{$\rho_{\xi_1,1}\otimes\rho_{\xi_1,2}$} gives rise to a $\mathrm{BosonSamp}(2)$ computer, while \mbox{$\rho_{\xi_1,1}\otimes\rho_{\xi_2,2}$} gives rise to two independent instances of $\mathrm{BosonSamp(1)}$ computers. Therefore, Eq. \ref{eq:expansion_simple} gives us a probabilistic mixture of two $\mathrm{BosonSamp}(2)$ computers and four $\mathrm{BosonSamp(1)}$ computers. Upon measuring the output statistics, we are classically sampling across these multiple instances of quantum Boson-sampling.

\begin{figure}[!htb]
\includegraphics[width=0.5\columnwidth]{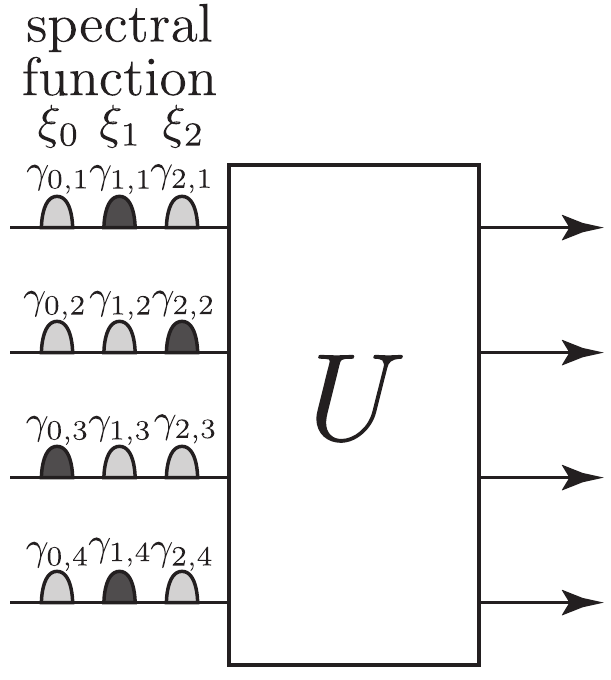}
\caption{An example instance vector \mbox{$v=(1,2,0,1)$} for an \mbox{$n=m=4$} Boson-sampling computer. There are three spectral basis states, the wave-packets shown in grey, each defined by some mode-function $\xi_i$, in this case localised temporal packets. For a particular $v$, the photons will occupy some combination of the allowed spectral basis functions, shown in dark grey. The total input state $\rho_\mathrm{in}$ is given by a summation over all allowed $v\in V$, weighted by products of the coefficients $\gamma_{i,j}$. In this example, $v$ gives rise to two $\mathrm{BosonSamp}(1)$ computers (in modes $\xi_0$ and $\xi_2$ respectively) and one $\mathrm{BosonSamp}(2)$ computer (in mode $\xi_1$). Thus, for $n_\mathrm{hard}=2$, we postulate that this instance $v$ is hard, which occurs with probability \mbox{$p(v)=\gamma_{1,1}\gamma_{2,2}\gamma_{0,3}\gamma_{1,4}$}.} \label{fig:mixed_input}
\end{figure}

In general, we may expand Eq. \ref{eq:input_mixed} as
\begin{equation} \label{eq:general_mixed_expansion}
\rho_\mathrm{in} = \sum_{v\in V} \left(\prod_{j=1}^n \gamma_{v_j,j} \cdot \bigotimes_{j=1}^n \rho_{\xi_{v_j},j} \right)
\end{equation}
where $V$ is the set of all vectors of length $n$, with integer indices spanning the support of the spectral basis, $v$ is one such vector, and $v_i$ is the $i$th element of that vector. $v$ can be interpreted as an instance of Boson-sampling, and $v_i$ as which spectral basis function the $i$th photon is in for that instance. An illustrative example is shown in Fig. \ref{fig:mixed_input}

We are interested in the situation where our Boson-sampling computer is classically hard to simulate. To formalise this, let us pick some threshold number of indistinguishable photons, $n_\mathrm{hard}$, upon which Boson-sampling is defined to be hard, and we assume $n_\mathrm{hard}\leq n$. Thus $\mathrm{BosonSamp}(n')$ is hard for $n'\geq n_\mathrm{hard}$. In our expression for the spectrally mixed input state we have a mixture of many different terms across a basis of different spectral distributions. Upon measurement we are classically sampling across different Boson-sampling problems. For classical hardness we desire that the probability we are sampling from a classically hard problem be above some threshold, $p_\mathrm{hard}>\epsilon$.

Let $\#(v)$ denote the largest number of repetitions of a single integer in the vector $v$. For example, \mbox{$\#(1,2,1,2)=2$} and \mbox{$\#(1,2,3,2,2)=3$}. Then $v$ denotes an instance of hard Boson-sampling when $\#(v)\geq n_\mathrm{hard}$. In Eq. \ref{eq:general_mixed_expansion} we are classically sampling across many Boson-sampling problems, one for each term in the summation. Specifically, the term associated with vector $v$ gives rise to a $\mathrm{BosonSamp}(\#(v))$ computer with probability \mbox{$p(v)=\prod_{j=1}^n \gamma_{v_j,j}$}, and some additional smaller Boson-sampling computers if $\#(v)<n$.

The probability that we sample from a hard instance of Boson-sampling is thus given by
\begin{eqnarray} \label{eq:p_hard_mixture}
p_\mathrm{hard} &=& \sum_{v \,|\,\#(v)\geq n_\mathrm{hard}} \prod_{j=1}^n \gamma_{v_j,j} \nonumber \\
&=& \sum_{v \,|\,\#(v)\geq n_\mathrm{hard}} p(v).
\end{eqnarray}

Let us consider the special case where all of the photons are spectral mixtures, but identical. In this case $\gamma_{i,j}$ is independent of $j$. First consider the limiting case where all the photons are identical and spectrally pure. In this case we can set $\xi_0=\psi$ and there will only be one non-zero coefficient in the spectral decomposition, i.e. $\gamma_i=\delta_{i,0}$. Then $p(v)=1$ when \mbox{$v=(0,\dots,0)$}, otherwise $p(v)=0$. If we consider the most restrictive case where $n_\mathrm{hard}=n$, then there is only one surviving term in the summation when \mbox{$v=(0,\dots,0)$}, for which $p(v)=1$, and thus $p_\mathrm{hard}=1$. Therefore, in the limiting case of spectrally pure, identical photons, our Boson-sampling computer is always implementing a hard algorithm, as is expected.

Next we consider the other limiting case where the photons are identical, but maximally spectrally mixed across $b$ spectral basis states. In this case \mbox{$p(v)=1/b^n\,\forall\, v$} and the single photon purity is $\mathcal{P}=1/b$. A closed-form expression for $p_\mathrm{hard}$ is challenging, but a lower bound is easily obtained,
\begin{equation} \label{eq:p_hard_mixed}
p_\mathrm{hard} \geq \sum_{k=n_\mathrm{hard}}^n \binom{n}{k} \mathcal{P}^k (1-\mathcal{P})^{n-k}.
\end{equation}

\begin{figure}[!htb]
\includegraphics[width=0.8\columnwidth]{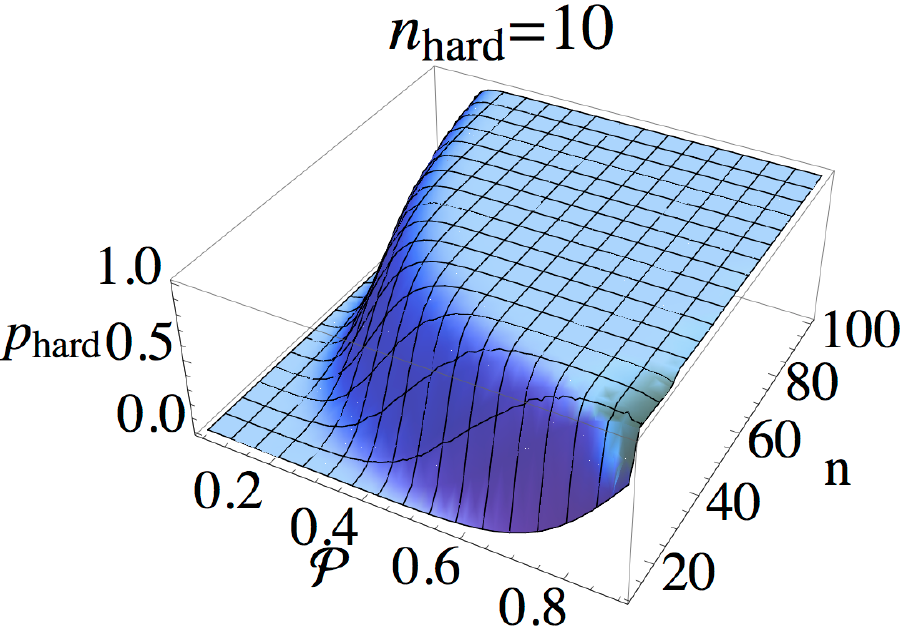}
\caption{(Colour online) Lower bound on $p_\mathrm{hard}$ for maximally spectrally mixed, but identical photons. For sufficiently large numbers of photons we can achieve computational hardness for arbitrarily low purities.} \label{fig:purity_3d}
\end{figure}

For maximally spectrally mixed states, the lower bound on $p_\mathrm{hard}$ is shown in Fig. \ref{fig:purity_3d}. The important feature of this plot is that even for highly impure photons we can achieve computational hardness if we use enough of them. Thus, for a given degree of hardness there is a tradeoff between the single photon purity and the required number of input photons. Specifically, as $n\to\infty$, the required purity $\mathcal{P}\to 0$. Thus, for identical but mixed photons of arbitrarily low purity we can always achieve computational hardness for sufficiently large systems.

The intuition behind this result is straightforward. For computational hardness we require that there exists some spectral mode within which reside at least $n_\mathrm{hard}$ photons. If the spectral basis states $\{\xi_i\}$ are being populated randomly, then clearly the probability of this occurring must asymptote to unity for large $n$.

This is an important observation for experimentalists, who are often limited by their photon source technology. For example, when preparing single photons by heralded parametric down-conversion (PDC), spectral correlations between the signal and idler photons will manifest themselves as spectral mixing in the heralded photon. For this reason much effort is invested into engineering PDC sources with separable spectral distributions. Additionally, uncertainty in the timing of the pump pulse will lead to temporal mixing. However, this result suggests that by scaling up the size of the system, such limitations may be overcome and computational hardness achieved nonetheless.

\subsection{Spectrally pure photons}

Next we turn our attention to spectrally pure photons, which in general have distinct spectral structures. The input state with arbitrary spectrally pure photons can be expressed
\begin{eqnarray}
\ket{\psi_\mathrm{in}} &=& \prod_{j=1}^n A_{\psi_j,j}^\dag \ket{0} \nonumber \\
&=& \prod_{j=1}^n \sum_i \lambda_{i,j} A_{\xi_i,j}^\dag \ket{0}.
\end{eqnarray}

In the spirit of Eq. \ref{eq:general_mixed_expansion} we may re-express our input state as
\begin{equation}
\ket{\psi_\mathrm{in}} = \sum_{v\in V} \left(\prod_{j=1}^n \lambda_{v_j,j} \cdot \prod_{j=1}^n A_{\xi_{v_j},j}^\dag \right) \ket{0},
\end{equation}
which is a sum over mutually orthogonal basis states. Again we desire that the total probability contributed by hard terms is above some threshold, so we define
\begin{eqnarray} \label{eq:p_hard_super}
p_\mathrm{hard} &=& \sum_{v \,|\,\#(v)\geq n_\mathrm{hard}} \left| \prod_{j=1}^n \lambda_{v_j,j}\right|^2 \nonumber \\
&=& \sum_{v \,|\,\#(v)\geq n_\mathrm{hard}} p(v).
\end{eqnarray}

Now we can consider two limiting cases. First, in the case of identical photons, applying the same reasoning as before, there is only one surviving term when $v=(0,\dots,0)$, and then $p_\mathrm{hard}=1$ as expected. The other limiting case is when all the photons are completely distinguishable and therefore evolve independently and do not interfere with one another. That is, each resides in a different basis state. In this case there is only one term in the superposition, corresponding to $v=(1,\dots,n)$, i.e. the $i$th photon resides entirely in $\xi_i$. Obviously the condition $\#(v)\geq n_\mathrm{hard}$ is not satisfied (except in the trivial case where $n_\mathrm{hard}=1$). Thus with completely distinguishable photons we are always implementing a classically easy computation ($n$ instances of $\mathrm{BosonSamp}(1)$), which is also expected.

We have considered the best- and worst-case scenarios for pure photons. We now consider the intermediate case, where the photons are pure but have some arbitrary degree of distinguishability. Deriving a completely general expression for this is prohibitive owing to the complicated combinatorics. Instead we will derive bounds on the operation of Boson-sampling where the worst-case photon distinguishability is known. That is, the minimum fidelity between any pair of photons is known, but we don't know the exact spectral decomposition for every photon. This is insightful as determining the fidelity between a pair of photons is relatively straightforward using simple interferometors, but performing full tomography of single photon states in the spectral degree of freedom is rather complex \cite{bib:Rohde06b} and to our knowledge has never been experimentally demonstrated.

Let us introduce a distinguishability parameter \mbox{$0\leq\alpha\leq 1$}, which captures the overlap between two photons' spectral decompositions. $\alpha$ is related to the fidelity as $\mathcal{F}=\alpha^2$. Let $\mathcal{F}_\mathrm{min}$ represent the worst-case fidelity between any two photons in our Boson-sampling system.

For a given value of $\mathcal{F}_\mathrm{min}$, the best-case operation of the system is when all the photons are identical, except for one which has overlap $\mathcal{F}_\mathrm{min}$ with the remainder. In this case the input state is
\begin{equation}
\ket{\psi_\mathrm{in}} = \left(\alpha A_{\xi_0,1}^\dag + \sqrt{1-\alpha^2} A_{\xi_1,1}^\dag\right) \prod_{i=2}^n A_{\xi_0,i}^\dag \ket{0}.
\end{equation}
Now there are two allowed state vectors, \mbox{$v=(0,0,\dots,0)$} and \mbox{$v=(1,0,\dots,0)$}, which occur with probabilities \mbox{$p(0,0,\dots,0)=\mathcal{F}_\mathrm{min}$} and \mbox{$p(1,0,\dots,0)=1-\mathcal{F}_\mathrm{min}$}. Since this represents the best-case scenario, substituting into Eq. \ref{eq:p_hard_super} we have
\begin{equation}
p_\mathrm{hard} \leq 1,
\end{equation}
for $n_\mathrm{hard}<n$.

\begin{figure}[!htb]
\includegraphics[width=0.8\columnwidth]{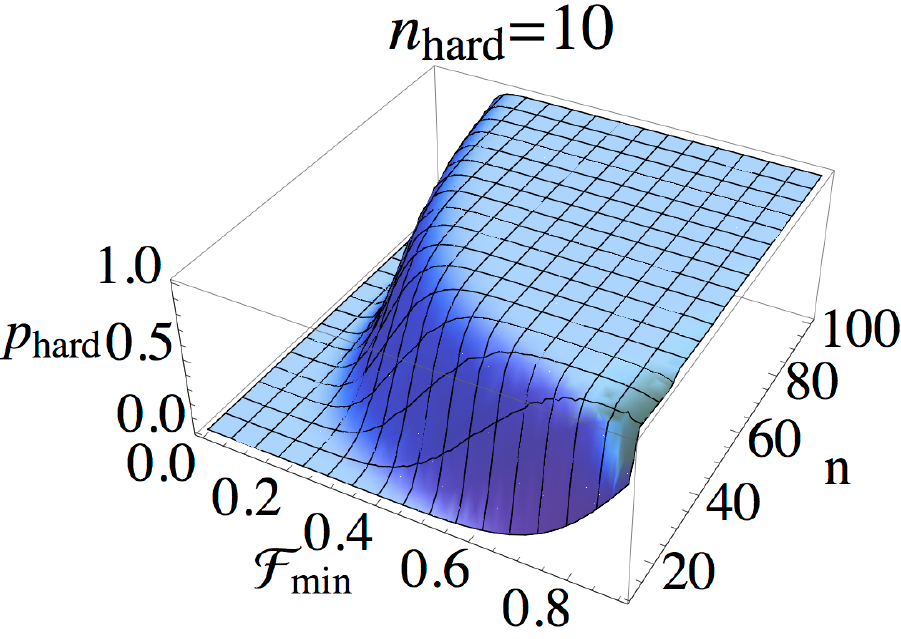}
\caption{(Colour online) Lower bound on $p_\mathrm{hard}$ against worst-case single photon fidelity and number of photons, for spectrally pure photons. For sufficiently large numbers of photons we can achieve computational hardness for arbitrarily low fidelities.} \label{fig:fidelity_3d}
\end{figure}

Alternately, for given $\mathcal{F}_\mathrm{min}$, the worst-case operation of the system is when all photons have overlap $\mathcal{F}_\mathrm{min}$ with all other photons. In this case we may write our input state as
\begin{equation}
\ket{\psi_\mathrm{in}} = \prod_{i=1}^n \left(\alpha A_{\xi_0,i}^\dag + \sqrt{1-\alpha^2} A_{\xi_i,i}^\dag\right) \ket{0}.
\end{equation}
Since this represents the worst-case scenario, substituting into Eq. \ref{eq:p_hard_super} we obtain
\begin{equation}
p_\mathrm{hard} \geq \sum_{k=n_\mathrm{hard}}^n \binom{n}{k} \mathcal{F}_\mathrm{min}^k (1-\mathcal{F}_\mathrm{min})^{n-k}.
\end{equation}
This lower bound on the hardness probability is illustrated in Fig. \ref{fig:fidelity_3d}, demonstrating that high $p_\mathrm{hard}$ may be achieved even with low worst-case fidelities, provided the system is sufficiently large. Again, this is a useful observation for experimentalists, who have inherent limitations in their photon fidelities. Indeed, as $n\to\infty$, the required $\mathcal{F}_\mathrm{min}\to 0$.

\begin{figure}[!htb]
\includegraphics[width=0.49\columnwidth]{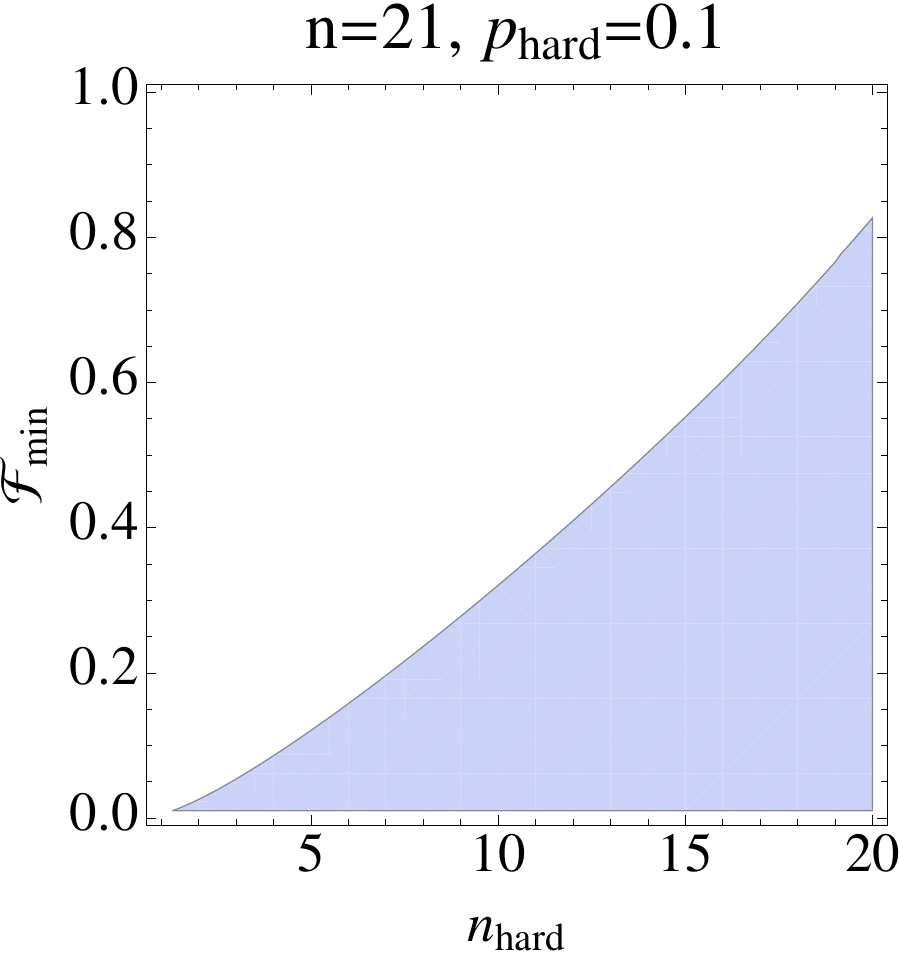}
\includegraphics[width=0.49\columnwidth]{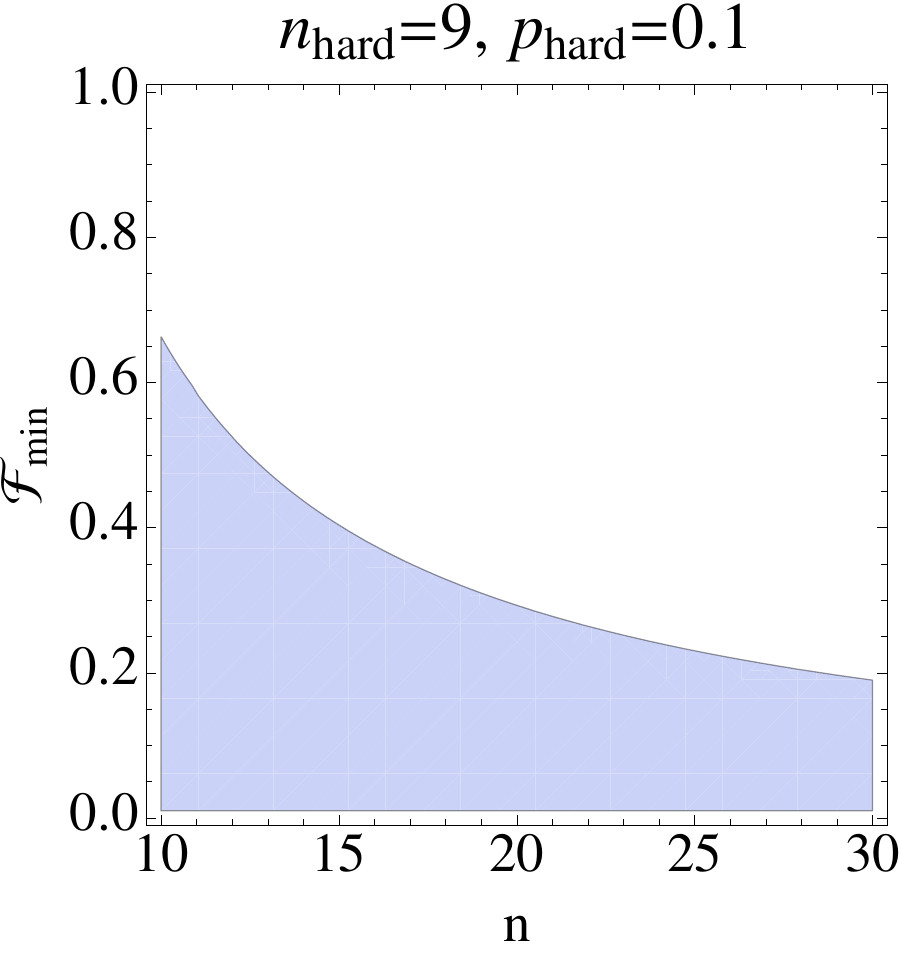}
\caption{Regions satisfying the inequality from Eq. \ref{eq:inequality}. $\mathcal{F}_\mathrm{min}$ is the minimum fidelity between any pair of photons in the system. (left) As our requirement for hardness increases (larger $n_\mathrm{hard}$), so does the upper bound on the required worst-case fidelity. (right) As the number of photons increases, the upper bound on the worst-case fidelity required to achieve a given degree of hardness decreases.} \label{fig:inequality}
\end{figure}

Our combined bound on $p_\mathrm{hard}$ is now
\begin{equation} \label{eq:inequality}
\sum_{k=n_\mathrm{hard}}^n \binom{n}{k} \mathcal{F}_\mathrm{min}^{k} (1-\mathcal{F}_\mathrm{min})^{n-k} \leq p_\mathrm{hard} \leq 1,
\end{equation}
for $n_\mathrm{hard}<n$. Fig. \ref{fig:inequality} illustrates examples of the regions satisfying this inequality.

\subsection{Interpretation of hardness}

Importantly, the regions satisfying $p_\mathrm{hard}>\epsilon$ in Eq. \ref{eq:p_hard_mixture} and Fig. \ref{fig:purity_3d} are not \emph{provably} hard. We cannot rule out the possibility there may exist some `tricks' enabling efficient computation for the mixed states in question (see Ref. \cite{bib:RohdeRalphAA12} for further discussion on this issue). For example, there are known quantum states with large average photon number (i.e. $\bar n\gg n_\mathrm{hard}$), but whose sampling problems are nonetheless classically easy (e.g. coherent states and other Gaussian states \cite{bib:Bartlett02sim, bib:Bartlett02bsim}). Thus, the regions within $p_\mathrm{hard}> \epsilon$ \emph{may} be classically hard, while the regions outside $p_\mathrm{hard}>\epsilon$ are \emph{definitely not} classically hard. That is, $p_\mathrm{hard}>\epsilon$ is definitely a necessary condition for computational hardness, but may not be a sufficient condition.

Similarly, in the case of pure states, Eq. \ref{eq:p_hard_super} and Fig. \ref{fig:fidelity_3d} do not specify provable hardness. Rather, we argue that the net contribution from terms which are individually hard is above some desired threshold. Of course there is in general interference between these different terms, so we cannot rule out the possibility that such interference effects make classical simulation easier.

Another important point is that the computational hardness of a Boson-sampling device is not just a function of the number of input photons $n$, but also of the unitary map $U$. Even with an ideal input state as per Eq. \ref{eq:ideal_input_state}, there are some unitary maps $U$ which are always computationally easy to simulate. The obvious examples are permutation matrices, which simply remap input modes to output modes without inducing any kind of interference effects. Obviously such a system can be trivially simulated for any input.

\subsection{Overcoming spectral effects}

In universal QC schemes, error correction and fault tolerant protocols can be employed to overcome the effects of errors and allow computation to proceed \cite{bib:NielsenChuang00}. However, in the limited Boson-sampling architecture there are no known error correction techniques. Thus fault tolerance may not be possible. One technique that is widely used in LOQC experiments to overcome mode-mismatch is to employ narrowband spectral filtering. That is, we employ frequency-resolving detectors (or detectors with a frequency filter) and post-select on events where the photons are within a narrow frequency range. This has the effect of projecting the photons onto a spectral structure whereby the different photons appear indistinguishable. While this technique is very useful and widely employed in elementary demonstrations, it is very limited since post-selection/filtering is equivalent to loss -- it discards a large part of the wave-packet, thereby reducing the detection probability. Thus, if there are $n$ photons in the system and the single photon efficiency is $p$, the probability of detecting all $n$ photons is $p^n$, which drops exponentially against $n$. In Boson-sampling we ideally wish to use a large number of photons to achieve complexity beyond classical capabilities. Thus, due to its unfavourable scaling properties, the spectral filtering technique is not a satisfactory approach. However, our results suggest that if computational hardness is the only objective, then error correction may not be necessary. Rather, we simply scale our system to have more modes and more photons.

\section{Conclusion}

We have considered the operation of the Boson-sampling model for linear optics quantum computation where the photons have arbitrary spectral structures, considering both the cases of spectrally pure and spectrally mixed photons. We derived analytic conditions for the relationship between the potential hardness of the computation and the spectral structure of the input photons.

We observed that with spectrally impure photons of arbitrarily low purity, computational hardness may be achieved by scaling up the size of the system. And for spectrally pure photons, the worst-case pairwise fidelity can be used to construct a lower bound on the hardness probability of the computer, and computational hardness may also be achieved for sufficiently large systems even with arbitrarily low fidelities.

Our results suggest the otherwise stringent requirements on photon indistinguishability in optical quantum computing schemes may be significantly relaxed at the expense of a larger system.

While we have not presented formal hardness proofs, our results build on those of Aaronson \& Arkhipov, and provide circumstantial evidence that Boson-sampling remains hard even with highly imperfect photon sources, provided we scale our systems sufficiently. Therefore, present-day limitations in photon engineering technology needn't prevent us from constructing devices with capabilities beyond those of classical computers, and demonstrating such devices may be realistic in the medium-term.

\begin{acknowledgments}
We thank Mike Hirschhorn, Sukhwinder Singh, Stephanie Wehner, Sean Seefried, Alexei Gilchrist, Dominic Berry, Nora Tischler, Alex Hayes, Daniel Horsley, Ben Toner, Catherine Greenhill and Danny Terno for assistance with a combinatorial problem that never got solved and doesn't appear in this paper. This research was conducted by the Australian Research Council Centre of Excellence for Engineered Quantum Systems (Project number CE110001013).
\end{acknowledgments}

\bibliography{paper}

\end{document}